\documentstyle[12pt,amsbsy]{article}
\newcommand{\bmv}[1]{{\boldsymbol #1}}
\begin{document}

\begin{center}
{\LARGE
Impurity effective mass in superfluid $^4$He}\\
\bigskip
{\em I.O.Vakarchuk, V.V.Babin \footnote{babin@ktf.franko.lviv.ua}.}\\
\bigskip
{\small
Ivan Franko Lviv State University\\
Department for Theoretical Physics\\
12 Drahomanov Str., 290005, Lviv\\  
Ukraine}
\end{center}
\begin{abstract}
 The system of bose liquid + impurity is considered.
 The energy spectrum as well as effective mass of impurity
 is calculated. For the case of the ${\rm ^3He}$ atom in
 superfluid ${\rm ^4He}$ numerical calculations are performed.

 {\bf keywords} {\it superfluid helium, impurity, effective mass.}

\end{abstract}

\section{Introduction}
   
   A couple of the most interesting questions is to be mentioned 
   within the problem of determining the impurity energy
   spectrum in superfluid ${\rm ^4He}$, namely the
   calculation of the impurity effective mass $M^{\ast}$ and
   revealing the spectrum character in the region of
   wave vectors corresponding to the reverse interatom
   distance. Within the frames of the thermodynamic
   approach the effective mass of ${\rm ^3He}$ was originally
   determined on the basis of experimental data for the heat
   capacity and the spin diffusion coefficient 
   by Bardeen J., Baym G., Pines D. \cite{BBP}.
   They found the value $M^{\ast}/M=2.34$ for the 
   impurity effective mass and the isolated atom $M$ mass ratio in the
   case of liquid ${\rm ^4He}$ equilibrium density. A
   microscopic theory of the energy spectrum for the ${\rm ^3He}$
   impurities in superfluid ${\rm ^4He}$ was
   independently developed in Ref.
   \cite{Bog,FD,SS_UPhZh,Woo1,Woo2,Vakarchuk}. Considering the
   case of a single ${\rm ^3He}$ atom, Devison T.B. and Feenberg
   E. \cite{FD} made use of the Brilluen-Wigner perturbation
   theory and calculated the impurity spectrum branch having
   chosen as zero approximation wave functions the wave
   functions of pure ${\rm ^4He}$ and isolated ${\rm ^3He}$
   atom. The authors determined the energy of replacement,
   the effective mass and relative change of liquid bulk
   coused by the replacement of ${\rm ^4He}$ with ${\rm ^3He}$. In
   order to obtain the numerical value of $M^{\ast}$, the
   liquid ${\rm ^4He}$ structure factor calculated
   theoretically by different authors was used. For example,
   using the data of \cite{MessWoo}, the value $M^{\ast}/M=1.81$
   is obtained. The same problems were considered by
   Slyusarev and Strzhemechny \cite{SS_UPhZh}; they used trial
   variation function of impure atom and the Brilluen-Wigner
   perturbation theory with the long wave estimate for the matrix
   elements of "impurity" and "impurity-phonon". An explicit
   expression for the impurity effective mass via the ${\rm ^4He}$
   structure factor as well as the numerical value
   $M^{\ast}/M=2.4$ on the basis of experimentally measured
   structure factor were given in this work too. The authors noted
   the essential non-square behaviour of the spectrum in the wave
   vector region near $3\AA^{-1}$. A similar expression for
   $M^{\ast}$ was obtained by Woo Tan  and Massey \cite{Woo1},
   however, by using the theoretical structure factor they
   obtained  $M^{\ast}/M=1.85$ to be closer to the value found
   in \cite{FD}.  The same authors in \cite{Woo2} have
   improved this result in $M^{\ast}/M=2.37$ by an
   evaluation of higher corrections to the effective mass of
   the interaction "impurity - phonon". In \cite{Owen}
   variational wave function suitable  only for the calculation
   of the effective mass is chosen.  With the help of this
   function which takes into account the backflow arising from
   driving the ${\rm ^3He}$ atom in the form proposed by Feynman
   and Cohen for pure  ${\rm ^4He}$ \cite{Feynman} the value
   $M^{\ast}/M=1.7$ is obtained. We would like to mention work
   \cite{GL} as well, in which the spectrum of the impurity
   and its dampings were investigated by means of the dynamic
   structure factor for small values of the wave vector
   $M^{\ast}/M=2.35$, as well as in the region of roton
   minimum ${\rm ^4He}$. "One-parameter" trial
   wave function without an application of the perturbations
   theory of Brilluen-Wigner is used in \cite{Vakarchuk} for 
   the calculation
   of the impurity branch of spectrum. For the effective mass
   two first corrections are explicitly calculated, each of
   which contains one sum over a wave vector more than the
   previous one. Using experimental data for the structure factor,
   in the case of the ${\rm ^3He}$ atom the value $M^{\ast}/M=1.73$
   is obtained.
   
   In the present work the calculation of the impurity 
   spectrum will be carried out on the basis of the polaron-type 
   Hamiltonian where liquid helium + impurity system is 
   simulated in the approximation of one sum over a wave 
   vector. We propose to pass in the Hamiltonian to other 
   variables so that impurity coordinates should be dropped 
   out, and further to develop the perturbation theory not 
   for the interaction "liquid - impurity", but for an 
   additional "anharmonic" term, which results from such a
   transformation of independent variables. In a zero 
   approximation of the approach the result for the spectrum 
   of an impurity of the second order of the perturbation 
   theory for a potential of the interaction "liquid - 
   impurity " is reconstituted, but without the assumption 
   of the smallness of the "liquid - impurity" interaction.
   Numerical estimates of an effective mass of the ${\rm^3He}$ 
   atom in superfluid ${\rm ^4He}$ on the basis of
   experimental values of the structural factor for liquid
   ${\rm ^4He}$ are also carried out.

\section{Notations}

   The object of our study is a model of the system 
   "superfluid ${\rm ^4He}$ + impurity", based on the
   following Hamiltonian:
   \begin{equation}
   \label{Hamiltonian_General}
      H=\frac{{\bmv P}^2}{2M}+H_{He^4}+H_{int}
   \end{equation}
   The first term in (\ref{Hamiltonian_General}) represents 
   a kinetic energy of the impurity atom (practically, it is the 
   ${\rm ^3He}$ atom), the mass of which we have denoted as $M$. We 
   take the Hamiltonian of superfluid helium in an
   approximation of noninteracting elementary  perturbations
   \begin{eqnarray*}
      H_{He^4}=E_{0}+\sum\limits_{{\bmv q}\neq 0}
      E_{q}b_{\bmv q}^{\dag}b_{\bmv q}
   \end{eqnarray*}
   with the Bogoliubov spectrum $E_q$ and energy 
   of a ground state $E_0$ \cite{BZ} :
   \begin{eqnarray*}
      &&E_0=\frac{N(N-1)}{2V}\nu_0-\frac{1}{4}
      \sum\limits_{{\bmv q}\neq 0} \frac{\hbar^2q^2}{2m}
      \left (
      \alpha_q-1
      \right )^2\\
      &&E_q=\frac{\hbar^2q^2}{2m}\alpha_q,\ \ \
      \alpha_q=\sqrt{1+\frac{2N}{V}\nu_q\left
      / \frac{\hbar^2q^2}{2m}\right .}\ \ ,
   \end{eqnarray*}
   where the letters $N$ and $V$ denote the number of
   particles of a liquid and the volume of the system
   respectively, $m$ is the mass of atom ${\rm ^4He}$ and
   $\nu_q$ is Fourie-image of a  pair potential of
   interparticle interaction in superfluid  ${\rm ^4He}$. The
   operators of creation-anihilation of the elementary 
   perturbations $b_{\bmv q}^{\dag}, b_{\bmv q}$ satisfy  the
   commutative relations of Bose
   \begin{eqnarray*}
      \left [ b_{\bmv q},b_{\bmv k}^{\dag} \right ] = \delta
      \left ( {\bmv q}-{\bmv k} \right ) \ .
   \end{eqnarray*}
   The part of Hamiltonian (\ref{Hamiltonian_General}) which 
   corresponds to the interaction between a liquid and impurity, 
   is expressed in the following form
   \begin{eqnarray*}
      H_{int}=\frac{N}{V}w_{0}+\frac{\sqrt N}{V}
      \sum\limits_{{\bmv q}\neq 0}
      \frac{w_q}{\sqrt \alpha_q}
      e^{i{\bmv q}{\bmv R}}
      \left (
      b_{-{\bmv q}}^{\dag}+b_{\bmv q}
      \right )\ .
   \end{eqnarray*}
   Here ${\bmv R}$ is a coordinate of an impurity, 
   $w_q$ is Fourie-image of a potential of interaction 
   between the ${\rm ^4He}$ atom and the impurity atom.

\section{Calculation of the impurity spectrum}

   In order to treat the problem
   it is convenient to pass from (\ref{Hamiltonian_General}) 
   to a unitary equivalent Hamiltonian $H_{\ast}$
   \begin{equation}
      H_{\ast}=UHU^{\dag}
   \end{equation}
   with the help of the transformation
   \begin{equation}
      \label{Transformation}
      U=exp
      \left (
      i\sum\limits_{{\bmv q}\neq 0}
      \left (
      {\bmv q}{\bmv R}
      \right )
      b_{\bmv q}^{\dag}b_{\bmv q}
      \right ), \ \ \
      U^{\dag}=U^{-1}\ .
   \end{equation}
   $U$  means the "translation" of the ${\rm ^4He}$ 
   atoms radious-vectors ${\bmv r}_1,...,{\bmv r}_N$ by
   $(-{\bmv R})$ that is a passage from the variables 
   $({\bmv r}_1,...,{\bmv r}_N,{\bmv R})$ to
   $({\bmv r}_1-{\bmv R},...,{\bmv r}_N-{\bmv R},{\bmv R})$.
   The explicit expression for $H_{\ast}$ is as follows
   \begin{eqnarray*}
      &&H_\ast=E_0^B+\frac{N}{V}w_0+
      \frac{1}{2M}
      \left (
      {\bmv P}-\sum\limits_{{\bmv q} \neq 0}
      \hbar{\bmv q}b_{\bmv q}^{\dag}b_{\bmv q}
      \right ) ^2+\\
      &&+\sum\limits_{{\bmv q} \neq 0}
      E_qb_{\bmv q}^{\dag}b_{\bmv q}+
      \frac{\sqrt N}{V}\sum\limits_{{\bmv q} \neq 0}
      \frac{w_q}{\sqrt \alpha_q}
      \left (
      b_{\bmv q}^{\dag}+b_{\bmv q}
      \right )
   \end{eqnarray*}
   Transformation (\ref{Transformation}) enables us to
   lose impurity coordinates, but, instead of that, results
   in the emergince of anharmonic terms.
   
   The operator of impulse ${\bmv P}$ of an impurity
   now can be considered as a c-number and
   one can identify a spectrum of impurity by the energy of a
   ground state which corresponds to $H_{\ast}$.

   The Hamiltonian $H_{\ast}$, as well as input $H$, cannot 
   be diagonalized exactly, therefore for further calculations 
   we use the theory of perturbations.
   It is natural in the next step to divide $H_{\ast}$
   into a problem that we supposes an exact solution,
   \begin{eqnarray*}
      &&H_\ast^0=\frac{N}{V}w_0+E_0+
      \frac{{\bmv P}^2}{2M}+\\
      &&+\sum\limits_{{\bmv q} \neq 0}
      \left \{
      \left (
      E_q+\frac{\hbar^2q^2}{2M}-\frac{\hbar}{M}{\bmv q}{\bmv P}
      \right )
      b_{\bmv q}^{\dag}b_{\bmv q}+
      \frac{\sqrt N}{V}\frac{w_q}{\sqrt \alpha_q}
      \left (
      b_{\bmv q}^{\dag}+b_{\bmv q}
      \right )
      \right \}
   \end{eqnarray*}
    and into perturbation
   \begin{eqnarray*}
      H_\ast^{int}=\frac{\hbar^2}{2M}
      \sum\limits_{{\bmv q}_1 \neq 0 }
      \sum\limits_{{\bmv q}_2 \neq 0}
      \left (
      {\bmv q}_1 {\bmv q}_2
      \right )
       b_{{\bmv q}_1}^{\dag} b_{{\bmv q}_2}^{\dag} 
       b_{{\bmv q}_1} b_{{\bmv q}_2},
   \end{eqnarray*} 
   (normal ordering of operators of creation-anihilation is
   suggested). But, by using an arbitrariness of this
   separation, we  introduce an additional parameter
   $\phi({\bmv q})$,  with the help of which we try to improve
   the approach of the  perturbations theory outcome to an
   exact solution. In other words, let's assume
   $$H_{\ast}={\widetilde H}_{\ast}^0+{\widetilde
   H}_{\ast}^{int}\ ,$$ where
   \begin{eqnarray*}
      {\widetilde H}_{\ast}^0&=&H_\ast^0+
      \sum\limits_{{\bmv q} \neq 0}
      \phi \left ( {\bmv q} \right )
      b_{\bmv q}^{\dag} b_{\bmv q}, \\
      {\widetilde H}_\ast^{int}&=&H_\ast^{int}-
      \sum\limits_{{\bmv q} \neq 0}
      \phi \left ( {\bmv q} \right )
      b_{\bmv q}^{\dag} b_{\bmv q} .
   \end{eqnarray*}
   In support of such a step it is possible to mention also 
   the argument that the small parameter of natural 
   origin does not exist in the input Hamiltonian $H$.
   
   We further develop the usual perturbation theory for an 
   energy of the ground state $H_{\ast}$ :
   \begin{eqnarray*}
      E=E^{(0)}+E^{(1)}+E^{(2)}+...
   \end{eqnarray*}
   
   In a zero approximation the system under consideration is 
   a set of  harmonic oscillators in an inhomogeneous 
   external field, therefore, without the assumption of the
   smallness of the liquid - impurity interaction, we obtain 
   \begin{eqnarray*}
      &&E^{(0)}=E_0+\frac{N}{V}w_0+
      \frac{{\bmv P}^2}{2M}-\\
      &&-
      \frac{N}{V^2}
      \sum\limits_{{\bmv q} \neq 0}
      \frac{w_q^2}
      {\alpha_q 
      \left (
      E_q+\frac{\hbar^2 q^2}{2M}-\frac{\hbar}{M}
      \left ( {\bmv q}{\bmv P} \right )+\phi \left ( {\bmv q} \right ) 
      \right )}.
   \end{eqnarray*}
   The first and the second correction to $E^{(0)}$
   have the following expression
   \begin{eqnarray*}
      &&E^{(1)}=-
      \frac{N}{V^2}
      \sum\limits_{{\bmv q} \neq 0}
      \phi \left ( {\bmv q} \right )
      \frac{w_q^2}
      {\alpha_q 
      \left (
      E_q+\frac{\hbar^2 q^2}{2M}-\frac{\hbar}{M}
      \left ( {\bmv q}{\bmv P} \right )+\phi \left ( {\bmv q} \right ) 
      \right ) ^2}+\\
      &&+\frac{\hbar^2}{2M}
      \left (
      \frac{N}{V^2}
      \sum\limits_{{\bmv q} \neq 0} {\bmv q}
      \frac{w_q^2}
      {\alpha_q 
      \left (
      E_q+\frac{\hbar^2 q^2}{2M}-\frac{\hbar}{M}
      \left ( {\bmv q}{\bmv P} \right )+\phi \left ( {\bmv q} \right ) 
      \right ) ^2}
      \right ) ^2 ,
   \end{eqnarray*}
   \begin{eqnarray*}
      &&E^{(2)}=-\frac{N}{V^2}
      \sum\limits_{{\bmv q} \neq 0}
      \frac{w_q^2}
      {\alpha_q 
      \left (
      E_q+\frac{\hbar^2 q^2}{2M}-\frac{\hbar}{M}
      \left ( {\bmv q}{\bmv P} \right )+\phi \left ( {\bmv q} \right ) 
      \right ) ^3} \times\\
      &&
      \times\left (
      \phi \left ( {\bmv q} \right )-
      \frac{\hbar^2}{M}
      \frac{N}{V^2}
      \sum\limits_{{\bmv k} \neq 0} 
      \left ({\bmv q}{\bmv k} \right )
      \frac{w_k^2}
      {\alpha_k 
      \left (
      E_k+\frac{\hbar^2 k^2}{2M}-\frac{\hbar}{M}
      \left ( {\bmv k}{\bmv P} \right )+\phi \left ( {\bmv k} \right ) 
      \right ) ^2}
      \right ) ^2+\\
      &&
      +\frac{N^2}{V^4}
      \sum\limits_{{\bmv q} \neq 0}
      \frac{\left ( \frac{\hbar^2 q^2}{2M}
      \right ) ^2 w_q^4}
      {\alpha_q^2
      \left (
      E_q+\frac{\hbar^2 q^2}{2M}-\frac{\hbar}{M}
      \left ( {\bmv q}{\bmv P} \right )+\phi \left ( {\bmv q} \right ) 
      \right ) ^5}-\\
      &&
      -\frac{N^2}{V^4}
      \sum\limits_{{\bmv q}_1 \neq 0 }
      \sum\limits_{{\bmv q}_2 \neq 0}
      \left (
      \frac{\hbar^2}{M} \left ( {\bmv q}_1 {\bmv q}_2  \right )
      \right ) ^2
      \frac{w_{q_1}^2}
      {
      \alpha_{q_1}
      \left (
      E_{q_1}+\frac{\hbar^2 {q_1}^2}{2M}-\frac{\hbar}{M}
      \left ( {\bmv q}_1{\bmv P} \right )+
      \phi \left ( {\bmv q}_1 \right ) \right ) ^2
      }\times\\
      &&
      \times\frac{w_{q_2}^2}
      {
      \alpha_{q_2}
      \left (
      E_{q_2}+\frac{\hbar^2 {q_2}^2}{2M}-\frac{\hbar}{M}
      \left ( {\bmv q}_2{\bmv P} \right )+
      \phi \left ( {\bmv q}_2 \right ) \right ) ^2
      }\times\\
      &&
      \times
      \frac{1}
      {
      E_{q_1}+\frac{\hbar^2 {q_1}^2}{2M}-\frac{\hbar}{M}
      \left ( {\bmv q}_1{\bmv P} \right )+
      \phi \left ( {\bmv q}_1 \right )+
      E_{q_2}+\frac{\hbar^2 {q_2}^2}{2M}-\frac{\hbar}{M}
      \left ( {\bmv q}_2{\bmv P} \right )+
      \phi \left ( {\bmv q}_2 \right )
      } .
   \end{eqnarray*}
   
   In order to fix the arbitrary function $\phi({\bmv q})$ 
   it is possible to use the demand of the minimization of
   energy as a functional of $\phi({\bmv q})$, but we did 
   not manage to solve the equations appearing
   in such an approach.
   We propose to require vanishing the first correction 
   to $E^{(0)}$, that is to determine $\phi({\bmv q})$ 
   from the condition $E^{(1)}=0$. Explicitly we have the
   equation :
   \begin{eqnarray*}
      &&-\frac{N}{V^2}
      \sum\limits_{{\bmv q} \neq 0}
      \phi \left ( {\bmv q} \right )
      \frac{w_q^2}
      {\alpha_q 
      \left (
      E_q+\frac{\hbar^2 q^2}{2M}-\frac{\hbar}{M}
      \left ( {\bmv q}{\bmv P} \right )+\phi \left ( {\bmv q} \right ) 
      \right ) ^2}+\\
      &&+\frac{\hbar^2}{2M}
      \left (
      \frac{N}{V^2}
      \sum\limits_{{\bmv q} \neq 0} {\bmv q}
      \frac{w_q^2}
      {\alpha_q 
      \left (
      E_q+\frac{\hbar^2 q^2}{2M}-\frac{\hbar}{M}
      \left ( {\bmv q}{\bmv P} \right )+\phi \left ( {\bmv q} \right ) 
      \right ) ^2}
      \right ) ^2 = 0 .
   \end{eqnarray*}
   The equation will be satisfied identically, if one puts
   $$\phi({\bmv q})=\frac{\hbar}{M}\left ({\bmv q}{\bmv x} \right )$$
   and imposes on the vector ${\bmv x}$ a condition
   \begin{equation}
      \label{Equation}
      {\bmv x}=\frac{\hbar}{2}\frac{N}{V^2}
      \sum\limits_{{\bmv q} \neq 0} 
      {\bmv q}\frac{w_q^2}{\alpha_q \left (
      E_q+\frac{\hbar^2q^2}{2M}-\frac{\hbar}{M}
      \left ( {\bmv q},{\bmv P-x} \right )
      \right ) ^2} .
   \end{equation}
   At ${\bmv P}=0$ we herefrom find ${\bmv x}=0$. For small
   ${\bmv P}$ let us assume that
   ${\bmv x}=\lambda{\bmv P}$, then, expanding the right-hand 
   side (\ref{Equation}) in the powers of ${\bmv P}$ we obtain the 
   following relation for the definition of $\lambda$
   \begin{eqnarray*}
     \lambda=\left ( 1-\lambda\right )\frac{\sigma_1}{2} + o(P^2)\ .
   \end{eqnarray*}
   Here for the some of simplicity we introduce
   \begin{eqnarray*}
      \sigma_1=\frac{4}{3}\frac{N}{V^2}
      \sum\limits_{{\bmv q}\neq 0}\frac{w_q^2
      \frac{\hbar^2q^2}{2M}}{\alpha_q
      \left (
      E_q+\frac{\hbar^2q^2}{2M}
      \right )^3}.
   \end{eqnarray*}
   Hence, within the accuracy of the square
   and higher  powers of impulse of the impurity, we have
   \begin{eqnarray*}
      \lambda=\frac{\sigma_1}{2+\sigma_1}.
   \end{eqnarray*}
   The usual theory of perturbations for $H_{\ast}^{int}$ corresponds 
   to the value $\lambda=0$.
   
\section{Effective mass of the impurity}

   The spectrum obtained above is in the region of small 
   impulses, as it should be, looks like
   $$E(P)=\varepsilon_0+\frac{P^2}{2M^{\ast}},$$
   where $M^{\ast}$ is the so-called effective mass of 
   the impurity atom, and the constant independent of impulse
   equals
   \begin{eqnarray*}
      &&\varepsilon_0 = E_0+\frac{N}{V}w_0-\frac{N}{V^2}
      \sum\limits_{{\bmv q}\neq 0}\frac{w_q^2}{\alpha_q
      \left ( E_q+\frac{\hbar^2q^2}{2M}
      \right )}+\\
      &&+\frac{N^2}{V^4}\sum\limits_{{\bmv q}\neq 0}
      \frac{\left ( \frac{\hbar^2q^2}{2M}\right ) ^2w_q^4}
      {\alpha_q^2
      \left (
      E_q+\frac{\hbar^2q^2}{2M}
      \right )^5}-\frac{N^2}{V^4}\sum\limits_{{\bmv q}_1\neq 0}
      \sum\limits_{{\bmv q}_2\neq 0}
      \left (
      \frac{\hbar^2}{M}\left ( {\bmv q}_1{\bmv q}_2 \right )
      \right ) ^2\times \\
      &&\times
      \frac{w_{q_1}^2}{\alpha_{q_1}
      \left (
      E_{q_1}+\frac{\hbar^2q_1^2}{2M}
      \right ) ^2 }
      \frac{w_{q_2}^2}{\alpha_{q_2}
      \left (
      E_{q_2}+\frac{\hbar^2q_2^2}{2M}
      \right ) ^2 }
      \frac{1}{
      \left (
      E_{q_1}+\frac{\hbar^2q_1^2}{2M}+
      E_{q_2}+\frac{\hbar^2q_2^2}{2M}
      \right ) }.
   \end{eqnarray*}
   The outcome of our calculations is such an expression for $M^{\ast}$
   \begin{equation}
      \label{EffMass}
      \frac{M}{M^{\ast}}=1-\sigma_1+\sigma_1^2-\sigma_1^3-\sigma_2+
      2\lambda\left ( \sigma_1^3+\sigma_2 \right )-
      \lambda^2\left ( \sigma_1^2+\sigma_1^3+\sigma_2 \right ) ,
   \end{equation}
   where one more notation is introduced
   \begin{eqnarray*}
      &&\sigma_2={\left ( \frac{4}{3}\frac{N}{V^2}\right )}^2
      \sum\limits_{{\bmv q}_1\neq 0}\sum\limits_{{\bmv q}_2\neq 0}
      \frac{\frac{\hbar^2q_1^2}{2M}\frac{\hbar^2q_2^2}{2M}
      w_{q_1}^2w_{q_2}^2}
      {\alpha_{q_1}\alpha_{q_2}
      \left (
      E_{q_1}+\frac{\hbar^2q_1^2}{2M}
      \right )^2
      \left (
      E_{q_2}+\frac{\hbar^2q_2^2}{2M}
      \right )^2}\times\\
      &&\times\frac{1}{\left (
      E_{q_1}+\frac{\hbar^2q_1^2}{2M}+
      E_{q_2}+\frac{\hbar^2q_2^2}{2M}
      \right )}\left \{
      \frac{\hbar^2q_1^2}{2M}\left [
      \frac{3}{\left ( E_{q_1}+\frac{\hbar^2q_1^2}{2M}\right ) ^2}+
      \right . \right . \\
      &&\left . \left .
      +\frac{2}{\left ( E_{q_1}+\frac{\hbar^2q_1^2}{2M} \right )
      \left ( E_{q_1}+\frac{\hbar^2q_1^2}{2M}+
      E_{q_2}+\frac{\hbar^2q_2^2}{2M}\right ) }+
      \right . \right . \\
      &&\left . \left .
      +\frac{1}{\left ( E_{q_1}+\frac{\hbar^2q_1^2}{2M}+
      E_{q_2}+\frac{\hbar^2q_2^2}{2M}\right ) ^2}
      \right ]
      +\frac{\hbar^2q_2^2}{2M}\left [
      \frac{3}{\left ( E_{q_2}+\frac{\hbar^2q_2^2}{2M}\right ) ^2}+
      \right . \right . \\
      &&\left . \left . +
      \frac{2}{\left ( E_{q_2}+\frac{\hbar^2q_2^2}{2M} \right )
      \left ( E_{q_1}+\frac{\hbar^2q_1^2}{2M}+
      E_{q_2}+\frac{\hbar^2q_2^2}{2M}\right ) }+
      \right . \right . \\
      &&\left . \left .
      +\frac{1}{\left ( E_{q_1}+\frac{\hbar^2q_1^2}{2M}+
      E_{q_2}+\frac{\hbar^2q_2^2}{2M}\right ) ^2}
      \right ]
      \right \} .
   \end{eqnarray*}

\section{Numerical calculations}

   We provide numerical computings of the effective mass of 
   the ${\rm ^3He}$ impurity  in ${\rm ^4He}$ at $T=0~K$. We
   assumed that  the potentials of interatomic interactions
   ${\rm ^3He-^4He}$  and ${\rm ^4He-^4He}$ are identical,
   that is $\nu_q=w_q$, and  further we expressed $\nu_q$
   through the ${\rm ^4He}$ structure factor with the help of
   relations  \cite{Bog}
   \begin{eqnarray*}
      S_q=\frac{1}{\alpha_q} .
   \end{eqnarray*}
   For evaluations we made use of measurements $S_q$
   \cite{SQ} continued to the point $T=0~K$ by means of 
   equalities \cite{Bog}
   \begin{eqnarray*}
      S_q(T=0)=S_q(T)\tanh\left [ \frac{E_q}{2T}\right ]\ .
   \end{eqnarray*}
   The density of ${\rm ^4He}$ was considered to be equal to 
   $\rho_{\rm ^4He}=0.02185\AA^{-3}$.
   The following values are obtained:
   \begin{eqnarray}
      &&\sigma_1=0.416\\
      &&\sigma_2=0.222
   \end{eqnarray}
   And, accordingly,
   \begin{eqnarray*}
      \frac{M^{\ast}}{M}=1.82 .
   \end{eqnarray*}
   The usual theory of perturbations $(\lambda=0)$ gives
   \begin{eqnarray*}
      \frac{M^{\ast}}{M}=2.15 .
   \end{eqnarray*}

\end{document}